\documentclass[preprint,journal]{vgtc}       

\ifpdf
  \pdfoutput=1\relax                   
  \usepackage{graphicx}                
  \DeclareGraphicsExtensions{.pdf,.png,.jpg,.jpeg} 
\else
  \usepackage{graphicx}                
  \DeclareGraphicsExtensions{.eps}     
\fi%

\graphicspath{{figures/}{pictures/}{images/}{./}} 

\usepackage{microtype}                 
\PassOptionsToPackage{warn}{textcomp}  
\usepackage{textcomp}                  
\usepackage{mathptmx}                  
\usepackage{times}                     
\usepackage{cite}                      
\usepackage{tabu}                      
\usepackage{booktabs}                  

\usepackage{xcolor}
\usepackage{enumitem}
\usepackage{fontawesome}
\usepackage{colortbl}

\usepackage{float}

\newcommand{\papertitle}{Striking a Balance: Reader Takeaways and Preferences when Integrating Text and Charts}

\newcommand{\pheading}[1]{\vspace{4px}\noindent\textbf{#1}}

\arrayrulecolor{gray} 

\newenvironment{tight_itemize}{\begin{itemize} \itemsep
-2.1pt}{\end{itemize}}

\usepackage{tabularx}
\usepackage{tabu}                      
\newcolumntype{P}[1]{>{\arraybackslash}p{#1}}

\usepackage{xurl}
\usepackage{changes}

\ieeedoi{10.1109/TVCG.2022.3209383}

\onlineid{1024}

\vgtccategory{Research}
\vgtcpapertype{Evaluation}

\title{\papertitle}

\author{Chase Stokes, Vidya Setlur (\textit{Member}), Bridget Cogley, Arvind Satyanarayan, and Marti A. Hearst}
\authorfooter{
\item
Chase Stokes is with UC Berkeley. E-mail: cstokes@ischool.berkeley.edu.
\item
 Vidya Setlur is with Tableau Research. E-mail: vsetlur@tableau.com.
\item
 Bridget Cogley is with Versalytix. E-mail: bcogley@versalytix.com.
 \item
 Arvind Satyanarayan is with MIT CSAIL. E-mail: arvindsatya@mit.edu.
 \item
 Marti Hearst is with UC Berkeley. E-mail: hearst@berkeley.edu.
}

\shortauthortitle{Stokes \MakeLowercase{\textit{et al.}}: \papertitle}

\abstract{ 
While visualizations are an effective way to represent insights about information, they rarely stand alone. When designing a visualization, text is often added to provide additional context and guidance for the reader. However, there is little experimental evidence to guide designers as to what is the right amount of text to show within a chart, what its qualitative properties should be, and where it should be placed.  Prior work also shows variation in personal preferences for charts versus textual representations. In this paper, we explore several research questions about the relative value of textual components of visualizations. 302 participants ranked univariate line charts containing varying amounts of text, ranging from no text (except for the axes) to a written paragraph with no visuals. Participants also described what information they could take away from line charts containing text with varying semantic content. We find that heavily annotated charts were not penalized. In fact, participants preferred the charts with the largest number of textual annotations over charts with fewer annotations or text alone. We also find effects of semantic content. For instance, the text that describes statistical or relational components of a chart leads to more takeaways referring to statistics or relational comparisons than text describing elemental or encoded components. Finally, we find different effects for the semantic levels based on the placement of the text on the chart; some kinds of information are best placed in the title, while others should be placed closer to the data. We compile these results into four chart design guidelines and discuss future implications for the combination of text and charts.
}


\CCScatlist{ 
 \CCScat{K.6.1}{Management of Computing and Information Systems}%
{Project and People Management}{Life Cycle};
 \CCScat{K.7.m}{The Computing Profession}{Miscellaneous}{Ethics}
}

\teaser{
   \centering
   \includegraphics[width=\linewidth]{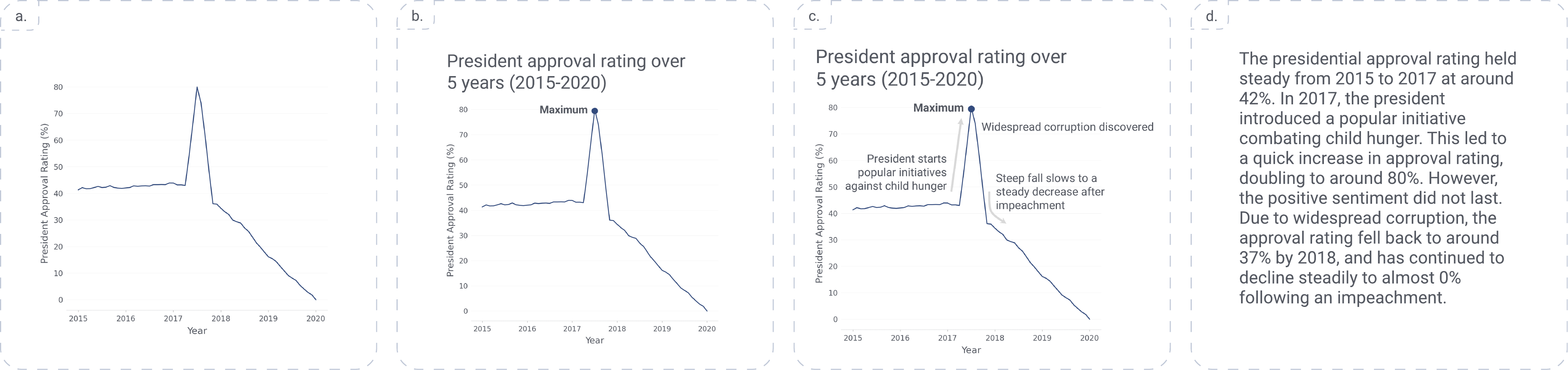}
   \caption{Information displays with varying amounts of visuals and text.  (a) Chart presented with no text (beyond axes and ticks), (b) Chart with a title and a single annotation, (c) Chart which displays a narrative or story around the data, annotated through text, and (d) A text-only version of the data, with the same story as displayed in (c). }
   \label{fig:teaser}
 }

\vgtcinsertpkg

\begin{document}


\firstsection{Introduction}
\maketitle

\firstsection{Introduction}

Information visualization has been defined as ``the use of computer-supported, interactive, visual representations of abstract data in order to amplify cognition'' \cite{card1999using}. 
In research, visualization is often contrasted with alternative forms of presentation such as tabular or written descriptions.  However, in reality, most charts are displayed with accompanying text in the form of titles, annotations, and captions. Designers include text with charts to convey the core message of the data, draw readers' attention to specific parts of the chart, and provide explanations that readers may be unlikely to identify on their own \cite{kosara2013storytelling}.  Appropriate text can provide additional context, potentially increasing the aesthetic appeal of a chart \cite{segel2010narrative}.

Textual descriptions have been shown to be  influential over the visual components of charts. Recent research has shown that the way a title is written can be the influencing factor on how a visualization is interpreted \cite{kong2018frames} and that the textual component can be the most memorable part of a visualization in a memory task \cite{borkin2015beyond}.

Although some recent research has acknowledged the important role that textual context plays in visualizations, there is currently little scientific understanding of the best qualities, quantities, and positions of textual content within charts and graphs. Additionally, information visualization studies often omit a text-only option when comparing charts \cite{stokesgive}, despite evidence that some people prefer text over charts in certain circumstances, such as in conversation with chatbots \cite{hearst2019would}.

The goal of this paper is to further our understanding of the role of text in visualizations and provide guidelines for both designers and automated systems. We examine questions around how much text people prefer on a chart, what kind of text is useful to show in terms of the semantics of that text, and how its placement affects the chart reader's understanding of the underlying data.

Text is often presented using different semantic content, ranging from descriptions of the overall data, to specific visual features, to additional context to support framing the chart \cite{lundgard2021accessible}. It is not currently known what content is best shown in the title versus in an annotation positioned near the data. It is also not known how much is too much text or if charts can stand alone with almost no annotation.  

This paper contributes a novel examination of the combination of text and charts in the context of visual analysis. By choosing a subset of a large design space, we examine relatively simple visualizations that offer a useful and nuanced examination of the rich space encompassing textual and visual communication. Our empirical studies find that readers prefer a combination of text and charts over  charts alone, and that a sizable minority prefer text without visualization. Our results show that the choice of semantic content and placement of annotations can influence what the readers take away from the chart and how likely they are to use the annotations in their takeaways. We provide a set of guidelines to authors for choosing the type of content and determining its placement for making the intended message most salient.

\maketitle

\section{Related Work}
\label{section:rw}

Our work exploring how readers integrate text with charts builds on prior work examining text for visualizations through alternative text and textual descriptions, empirical results indicating the influence of text on the reader, and considerations from systems and applications of these two modalities to visual analysis and data storytelling.

\subsection{Accessibility (Blind and Low Vision)}

Text is a primary means for making visualizations accessible to people who are blind or have low vision (BLV). The W3C Web Accessibility Initiative (WAI), visualization practitioners, and publishers of accessible media have produced a variety of guidelines for what these textual representations (referred to as alternative text or alt text) should seek to convey, including the chart type, axes and labels, and summaries of the overall trends \cite{w3c_wai_2019, cesal_writing_2020, turner2018benetech}. Building on these guidelines, Jung et al. surveyed and interviewed BLV people to find that alt text is often the only interface to the depicted data; poor quality text can increase inequities in information access \cite{jung2021communicating}. The researchers further recommended that alt text should be structured to ground the textual representation in the underlying data rather than the visualized elements to reduce readers' cognitive burden. 

Most related to our work is a recent paper by Lundgard \& Satyanarayan that began to formalize the semantic content of textual descriptions of charts \cite{lundgard2021accessible}. They developed a conceptual model that spans four levels of semantic content: enumerating visualization construction properties (e.g., marks and encodings); reporting statistical concepts and relations (e.g., extrema and correlations); identifying perceptual and cognitive phenomena (e.g., complex trends and patterns); and elucidating domain-specific insights (e.g., social and political context). 

In studies with sighted and blind participants, the researchers found that these reader groups differ significantly in which semantic content they ranked as most useful. While sighted participants generally preferred descriptions that offered high-level explanations and domain-specific context, blind readers ranked this content lower and instead favored either mid-level descriptions of statistical features or perceptual trends, and expressed bi-modal preferences for low-level details about the chart's construction. Our work expands the study of the impact of these semantic levels.

\subsection{Role of Text in Visualizations}

Stokes and Hearst advocated for text to be considered co-equal to visualization, calling for researchers to devote attention to concerns of readability and integration between these two modalities \cite{stokesgive}. In this spirit, there is a growing body of work that explores the role that text plays in visual analysis and its importance for conveying key messages to the reader \cite{ottley2019curious}.

For instance, through eye-tracking studies, Borkin et al. \cite{borkin2015beyond} found that participants fixated on and were more likely to recall the textual content of and around visualizations (including titles, labels, paragraphs, etc.). Kong et al. \cite{kong2018frames,kong2019trust} evaluated how titles can impact the perceived message of a visualization to find that participants were more likely to recall information conveyed by slanted framings (e.g., emphasizing only part of a chart's message) than the chart's visuals. Troublingly, participants only flagged the slanted framing if it was especially egregious (e.g., offering a contradictory message to the chart); more subtle framings escaped unnoticed despite shaping participants' takeaways. 

In contrast to these results, Kim et al. found that readers are perhaps able to mediate misalignments between text and visualization \cite{kim2021towards}. Through crowdsourced studies, they found that when both the chart and text described a high-prominence feature, readers treated the doubly emphasized high-prominence feature as the takeaway. When the text described a low-prominence chart feature, readers relied on the chart and usually reported a higher-prominence feature as the takeaway.

Work by Mayer in multimedia communication provided rigorous empirical support for fifteen theory-based effects of how people learn from the combination of words and images \cite{mayer2020multimedia}.  One of these was the ``multimedia effect'', the hypothesis that people can learn more deeply from the combination of words and images together than either alone. This work and other studies have shown that text placed in spatial proximity to explanatory images can reduce cognitive load \cite{holsanova2005tracing, holsanova2009reading, zhao2014eye}. 

However, in the context of information visualization, this combination of words and images is not always \textit{preferred} by a reader. Hearst \& Tory examined participant preferences for visualizations in conversation with chatbots \cite{hearst2019would}. They found a large split in participant preferences, with almost half of the readers preferring not to see charts. For those who \textit{did} prefer to see charts, they also preferred to view additional contextual data in the chart. The role of overall preference for textual or visual communication has been shown to influence reader preferences for a combination of the two and therefore, should be accounted for in this study as well.

\subsection{Visualization + Text for Analysis and Storytelling}

Researchers are increasingly exploring how to leverage the affordances of natural language text during visual analysis \cite{srinivasan2017natural}. For instance, systems like Articulate \cite{sun2010articulate}, DataTone \cite{gao2015datatone}, and Eviza \cite{setlur2016eviza} demonstrate that the familiarity of natural language can provide a shallower on-ramp to visual analysis, and the ambiguity of these statements can be used to rapidly suggest candidate visualizations. Subsequent work also explored how pragmatics \cite{hoque2017applying}, and question-answering pipelines \cite{dhamdhere2017analyza, kim2020answering} can aid in a more conversational interaction model. 

Besides serving as input or output for visual analysis, systems have begun to incorporate natural language-generated text along with their visualization responses to help describe key insights to the user. For example, a variety of research systems \cite{mittal1998describing, srinivasan2018augmenting, kanthara2022chart, chen2020figure, qian2021generating} and  tools such as Tableau's Summary Card \cite{tableausummary}, 
and Power BI \cite{powerbi} produce natural language captions that summarize statistics and trends depicted by the chart.

Data storytelling weaves data, textual narratives, and visuals together for sharing and communicating insights that are more memorable, persuasive, and engaging than statistics alone \cite{segel2010narrative,kosara2013storytelling, lee2015more}. With the growing recognition of using both text and charts for data storytelling, there is an added challenge of helping readers synthesize information from both these pieces of information in documents, especially when they are spatially separated.  

Systems like Kori \cite{latif2019authoring, latif2021kori} and VizFlow \cite{sultanum2021leveraging} provide explicit linking strategies between the two modalities following design patterns that have emerged for data storytelling, narrative sequencing, and rhetoric \cite{hullman2013deeper, hullman2011visualization}. While these systems demonstrate the utility of designing text and charts together, it is still unclear how to balance between the two modalities (e.g., is too much or too little text still useful?). Our work probes these research questions and provides design implications for future systems for jointly authoring text and visualization.

\begin{figure*}[t!]
 \centering
 \includegraphics[width=\linewidth]{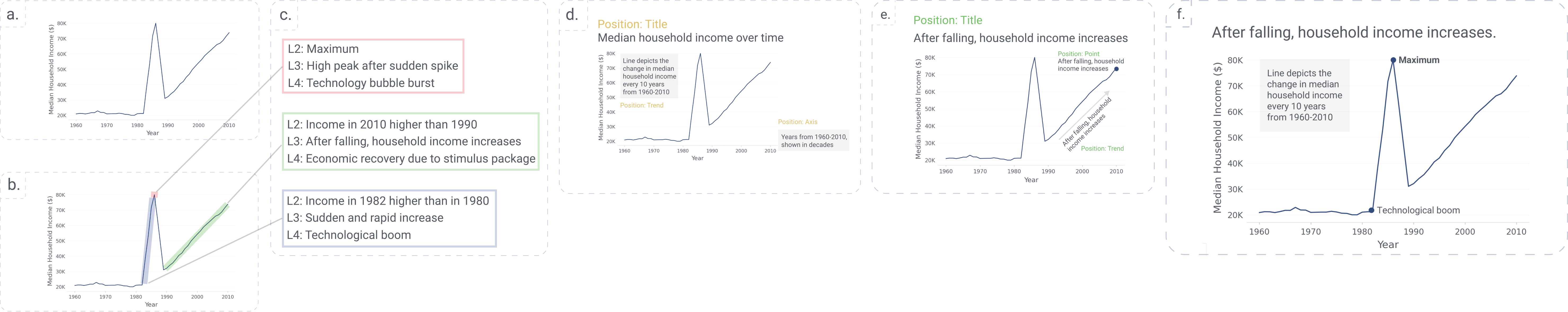}
 \caption{Stimuli creation process. Annotations were written and placed based on the components of the chart, which were identified as visually salient using the method from Kim, Setlur, \& Agrawala \cite{kim2021towards}. The initial chart is shown in (a). In (b), red indicates the most visually salient feature, green the second, and blue the third. In (c), paper authors created potential annotations for each salient region at each level L2--L4. L1 cannot refer to a part of the chart identified as salient, but as shown in (d), it can refer to different areas of the chart (axis, trend, and title). A `trend' placement referred to an annotation that describes the line as a whole. L2--L4 can each be placed in the same positions, as shown in (e); they can be expressed in a point, a trend, or as the title. In step (f), an expert designer adjusted the fine details to produce a chart with a realistic layout.  }
 \label{fig:stimuli_creation}
\end{figure*}

\section{Study}
\label{section:study}

In this study, we utilize the framework from Lundgard \& Satyanarayan \cite{lundgard2021accessible} which describes four semantic levels: 

\begin{tight_itemize}
\item \textbf{Semantic Level 1 (L1)}: The first level consists of elemental or encoded aspects of the chart, such as the overall topic or a description of the content of an axis. For example, in Figure \ref{fig:teaser}, \textit{``President approval rating over 5 years (2015-2020).''}
\item \textbf{Semantic Level 2 (L2)}: The second level consists of statistical or relational components, such as a comparison between two points or identification of extrema. For example, in Figure \ref{fig:teaser}, \textit{``Maximum''}.
\item \textbf{Semantic Level 3 (L3)}: The third level describes perceptual or cognitive aspects, such as an overall pattern or changes in trend. For example, in Figure \ref{fig:teaser}, \textit{``Steep fall slows to a steady decrease.''}
\item \textbf{Semantic Level 4 (L4)}: The fourth level provides external context to the chart, such as past events which affect the topic depicted. For example, in Figure \ref{fig:teaser}, \textit{``President starts popular initiatives against child hunger''}.
\end{tight_itemize}

\noindent
We explore three primary research questions and their corresponding hypotheses, using these four semantic levels as a way to examine the content of an annotation for RQ2 and RQ3:

\noindent \textbf{RQ1}: What are readers' preferences when viewing charts with different amounts of text? To investigate this question, participants ranked charts with varying amounts of text, including an all-text paragraph. We hypothesized the following: 

\begin{tight_itemize}
\item \textbf{H1a}: Readers with an overall preference for visual information will prefer charts with annotations compared to the all-text variant.
\item \textbf{H1b}: Readers with an overall preference for textual information will prefer the all-text variant compared to charts with little or no text.
\item \textbf{H1c}: Readers with an overall preference for visual information will prefer charts with fewer annotations.
\item \textbf{H1d}: Readers with an overall preference for textual information will prefer charts with a greater number of annotations.
\end{tight_itemize}

\begin{figure}[t]
 \centering
 \includegraphics[width=\linewidth]{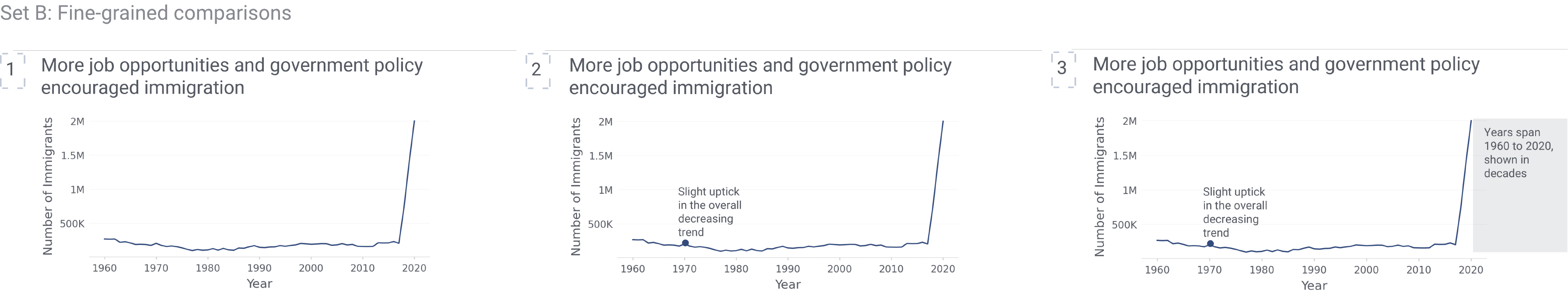}
 \caption{Example stimuli for ranking set B (fine-grained comparisons). An example of ranking set A (`extremes') appears in Figure~\ref{fig:teaser}. 
 }
 \label{fig:ranking_sets}
\end{figure}

In investigating these hypotheses, we created two sets of ranking stimuli. The first encompassed the `extremes' within the range from text to visual. This included a chart with no text (except axes) to a text paragraph with no chart (see Figure~\ref{fig:teaser}). The second set of stimuli examined more fine-grained differences, ranging from a chart with only a title to a chart with a title and two annotations (see Figure \ref{fig:ranking_sets}).

\medskip

\noindent \textbf{RQ2}: How do the different semantic levels of text affect the type of information included in readers' takeaways? Further, when do readers include information in their takeaways that are not found either in the chart or the text? Which semantic levels of text do readers rely on when they form their takeaways? 

Hypotheses were as follows:
\begin{tight_itemize}
\item \textbf{H2a}: Readers' takeaways will be more likely to contain a given semantic level if the text accompanying the chart also contains that semantic level. This will be true for all semantic levels.
\item \textbf{H2b}: Readers will be more likely to include information in their takeaways that is not found in the chart or text when L1 is present in the chart.
\item \textbf{H2c}: Readers will self-report as relying less on text when the text contains L1.
\item \textbf{H2d}: Readers' will self-report as relying more on text when the text contains L4.
\item \textbf{H2e}: Readers with an overall preference for textual information will rely more on text within a chart. 
\end{tight_itemize}

The feature mentioned in captions influences the amount of reader takeaways focused on a particular feature. This effect is greater when the caption includes external information (i.e., L4) \cite{kim2017explaining}. L1 content, on the other hand, provides the most basic information about a chart. These components motivate hypotheses H2a-H2d. Since there is evidence that readers with a lower preference for charts overall prefer to receive information through text \cite{hearst2019would}, we also hypothesize that they will rely more on the textual information (H2e).

\medskip

\noindent \textbf{RQ3}: How does the placement of text affect how the reader integrates the text with the chart? Our hypothesis was:
\begin{tight_itemize}
\item \textbf{H3}: Readers' takeaways will be most likely to contain a given semantic level if the text containing that semantic level is positioned as a title.
\end{tight_itemize}

When readers view a chart, they first focus on the title before moving to other textual components or the data\cite{borkin2015beyond}. We hypothesized this position would then be the most likely for the reader to use in a takeaway.

To investigate research questions 2 and 3, participants were asked to inspect a chart with varying annotations at varying semantic levels and then provide up to three takeaways. To test these hypotheses, we designed the stimuli to systematically vary the content and location of the textual annotations. Charts displayed in this section contained either a title alone or a title with a single annotation. The semantic levels of the text content varied. See Section \ref{section:stimuli} for further information on development of the stimuli. 

\subsection{Participants}

To determine the proper sample size for this study, we conducted a power analysis with G*Power \cite{faul2007g}. Based on a logistic regression with an odds ratio of 1.5 and a desired power of 0.8, the proper sample size (post exclusions) was 297, with an alpha level of 0.05. This sample size also accommodates the amount recommended from an additional power analysis on a Friedman's test.

Based on this information, we recruited 512 participants from Amazon Mechanical Turk. In order to qualify for the study, participants were required to be located in the United States and to have a 95\% acceptance rate on previous tasks. We also required that they be fluent in English and complete the survey on a desktop or laptop device. We compensated participants at a rate of \$15 per hour. This resulted in a \$3.50 compensation for a 14-minute survey, which was raised to \$4.00, due to a longer survey time than estimated. We employed comprehension checks in the first block of the survey, described further in Section \ref{section:method}. 
Participants were not permitted to take the survey multiple times. 

After excluding participants who failed the comprehension checks as well as any who provided extremely low-quality responses (e.g., takeaways that were either unintelligible or off-topic), we ended up with 302 participants. On average, participants were 35-44, and most had completed a 4-year degree. On a scale from 1 (overall preference for textual communication) and 6 (overall preference for visual communication) \cite{garcia2016measuring}, participants had an average score of 4.06 across the measures used (details in Section \ref{section:method}).

\subsection{Stimuli}
\label{section:stimuli}
\begin{figure}[ht]
 \centering
 \includegraphics[width=\linewidth]{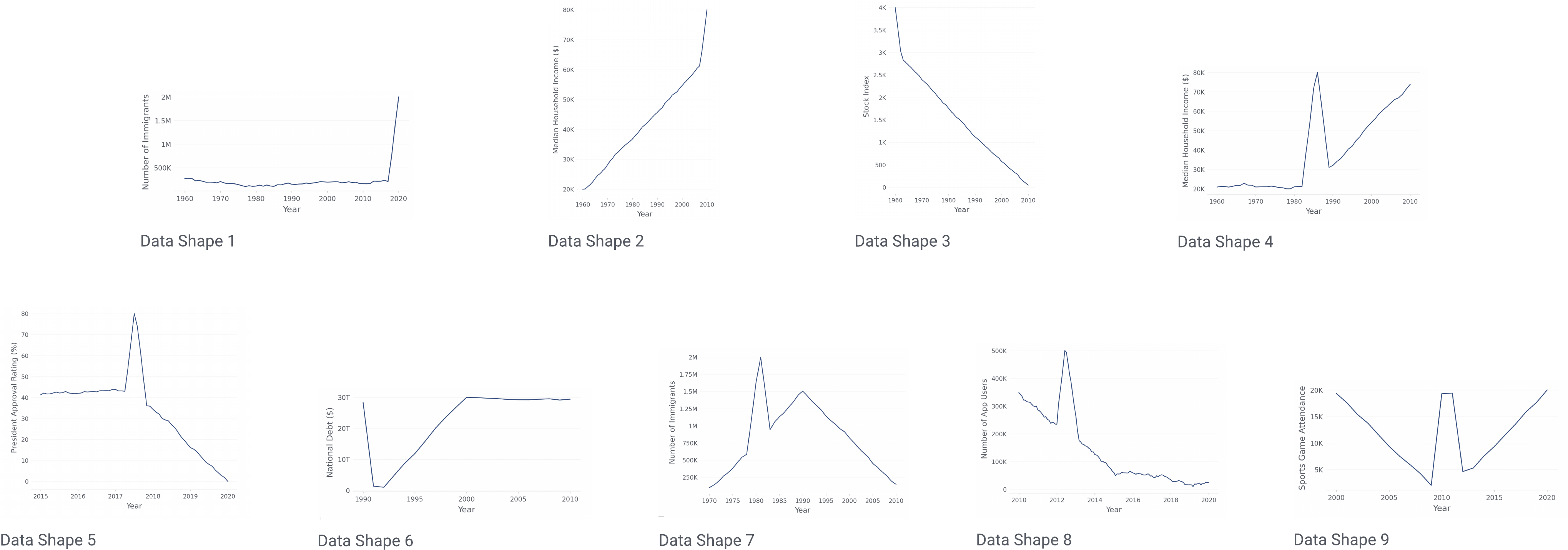}
 \caption{Nine univariate charts selected for this study, with at most two trends (up, down, or flat) and a single feature in the positive direction.}
 \label{fig:stimuli_set}
\end{figure}

All stimuli used in this study were univariate line charts, adapted from Kim et al. \cite{kim2021towards}. We considered univariate line charts for our work because they are among the most common basic charts, and because these charts contain temporal features that can be easily annotated, making them useful for exploring our hypotheses. A subset of nine charts from the original paper was selected to keep the overall design space tractable. These charts contained at most two trends (up, down, or flat), with a single feature in the positive direction. The full set of charts selected can be seen in Figure \ref{fig:stimuli_set}. This design space ensured relatively realistic global shapes with sufficient variation in the stimuli while maintaining enough blank space on the chart in which to situate textual and visual annotations.

We  added aesthetic design updates to the charts, including  lightening the gridlines and axis ticks. The data line itself was also darkened. These aesthetic changes were completed to align the stimuli to best practices as defined by practitioner guidelines \cite{knaflic_2017,berinato2016good,few2004show}, therefore imitating real-world professional charts. Advice and design style was provided by a paper author who is an expert visualization designer. To provide context to the charts, we labeled the x-axis with time unit values implying that the chart represents a time series. Specifically, we selected the start and end of the x-axis from the set of years {1900, 1910, 1920,..., 2020}. To label the y-axis, we randomly chose a set of domains for the y-axis and its value range from the MassVis dataset \cite{borkin:2013}.

For each chart in the stimuli set, we also included an all-text variant, as advocated for by Stokes \& Hearst \cite{stokesgive}. Three authors created textual descriptions of the chart, incorporating the additional information chosen to accompany the chart. These three versions were then discussed and synthesized into a single paragraph with feedback from all authors.

\subsubsection{Annotation content}

We wrote four possible textual annotations to be added to each chart. Each of the four annotations per chart contained information from one of the semantic levels outlined by Lundgard \& Satyanarayan \cite{lundgard2021accessible} as described at the beginning of this section.

These annotations were written by two authors 
to ensure that the semantic content was consistent with the descriptions for each of the four semantic levels. Annotations were not computationally generated; especially for L3 and L4, we wanted to ensure that the annotations reflected ``natural sounding" language, as required by the definitions of those levels. 

The annotation content was written based on the visually salient features in the line charts, as shown in Figure \ref{fig:stimuli_creation}. Additionally, the external context of L4 annotations was created with reference to Wikipedia articles pertaining to the topic of each chart to create plausible commentary \cite{wikipedia2004wikipedia}. Identification of these visually salient features was computed using crowdsourced data from Kim et al. \cite{kim2021towards}. 

Saliency mapping utilized mouse input information for bounding boxes that participants from \cite{kim2021towards} drew around features that they identified as the top three visually prominent features in the line charts. Given that the stimuli were time-series line charts, we aggregated all of the feature bounding boxes by projecting each box onto the x-axis containing time values, to form a 1D interval. We then weighted each interval inversely proportional to the ranking provided by the crowdsourced data. We then multiplied the weight assigned to each interval by a Gaussian factor centered at the interval and with standard deviation set to half the width of the interval. Summing all of the Gaussian weighted intervals, we obtained the top three visually salient regions for each line chart stimulus.

The visual salience of each semantic level was counterbalanced between the nine selected charts, such that each semantic level had the same average visual salience across the stimuli set. However, L1 did not refer to any particular point or area within the data and so could not be mapped to visually salient features. This counterbalancing was only completed for L2- L4. More detail about the stimuli creation process is shown in Figure \ref{fig:stimuli_creation}.
 
\subsubsection{Annotation appearance and position}

Annotations were added to the line charts with the aforementioned visually salient features.
The progression from a blank chart through the full set of possible annotations can be seen in Figure \ref{fig:stimuli_creation}. Design practices include standard conventions \cite{few2004show, knaflic2015storytelling, tufte1985visual, wong2010wall} to prioritize focus on the line. The line color was selected to blend in with the blue used elsewhere in the survey instrument without creating a visual artifact for users with astigmatism \cite{locke_2021}. All text met at least AA guidelines under WCAG 2.0 testing \cite{w3c_wcag}. The placement of the annotations was guided by the following design guidelines.

\begin{tight_itemize}
\item \textbf{Legibility}: Avoid adding annotations too close to one another or overlapping \cite{kirk2016data,  wong2010wall}.
\item \textbf{Visual language}: Add banding to depict time ranges, show arrows for trends, and dots for point-in-time references.
\item \textbf{Spatial agreement}: The label should follow the pattern of the data and be close to the referent. 
\end{tight_itemize}

The overall placement of the annotations was counterbalanced across the set and between the four semantic levels. For example, each semantic level was positioned in the title or subtitle position three times. These stimuli and their annotations went through several rounds of iteration through the processes in Figure \ref{fig:stimuli_creation}. 

An additional chart variant was created to convey the same story as the all-text variant, to preserve dynamic equivalence \cite{metzger_1999}. Any additional information provided in the all-text variant, such as a current event or emphasized feature, was added in the form of an annotation.

\subsection{Method}
\label{section:method}

\begin{figure*}[ht]
 \centering
 \includegraphics[width=\linewidth]{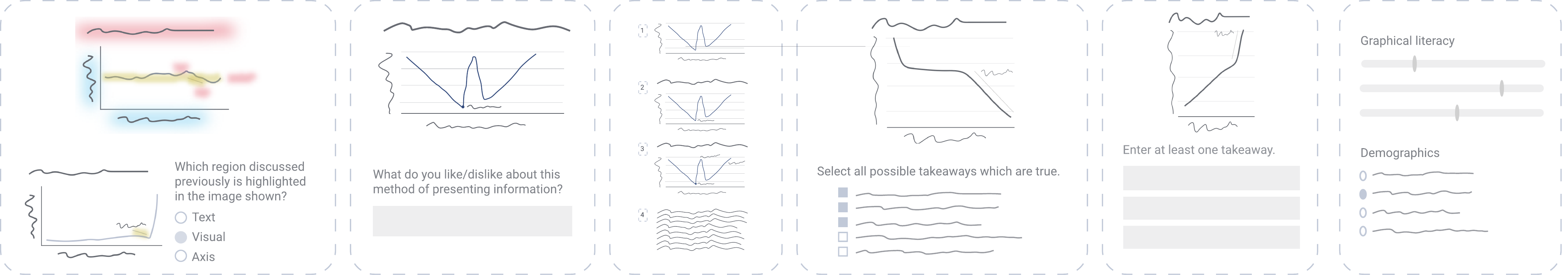}
 \caption{Participants completed a survey with the six sections above: terminology introduction, individual viewing of stimuli, ranking of the full stimuli set, filter task, entering takeaways, graphical literacy, and demographics. Individual viewing and ranking were completed for both ranking sets.}
 \label{fig:survey_flow}
\end{figure*}

Participants completed a survey with five main sections. The full survey flow can be seen in Figure \ref{fig:survey_flow}. First, participants were introduced to the study and the terminology used throughout the survey. They were instructed in how to apply the terms \emph{axis}, \emph{visual}, and \emph{text}, with \emph{text} deconstructed into \emph{title} and \emph{annotation}. They were then asked to answer 1-2 questions to check for understanding and were provided with an illustration of the terminology after each question. Those who answered the first question correctly were not asked the second question. If both comprehension checks were missed, the survey ended.

After this introduction, participants went on to complete the first set of ranking tasks. They were instructed: ``In the next section, you will rank two sets of images. During this section, keep in mind the following instructions. Rank the images in the order you would prefer to encounter or see them. You can rank by dragging and dropping the subsequent images into a rank.'' The ranking tasks within this survey were designed to elicit user preferences for the amount of text paired with a chart. Prior to completing the ranking, participants viewed each stimulus individually and answered a free-response question about the qualities of the information presentation which they liked and those which they disliked.

The stimuli for the first set of charts (Set A) consisted of four items: a chart with only axis labels and tick marks (no-text), a chart with a title and an annotation (title-annotation1), a chart with a story or narrative about the data annotated (annotation+), and a paragraph of text which described the chart shown (all-text). In set A, we aimed to capture preferences for the ‘extremes’ between an (almost) entirely visual presentation of information and an entirely textual presentation. The second set of charts (Set B) consisted of three charts: one with only a title (title-only), title-annotation1, and a chart with a title and two annotations (title-annotation2). With this set, we examined a more detailed variation in the number of annotations, progressively increasing it within the set. Specific chart combinations within sets were also selected to best counterbalance between annotation position and content. 

Following the ranking task, participants went on to complete a second comprehension check, which also provided them with some training on our definition of `takeaway'. Takeaways were defined as ``a key fact, point, or idea to be remembered after viewing the chart.'' 
In this filter task, participants were shown a chart and five example takeaways that could be drawn from the chart. Three of the five were true statements about the chart, while two were false. Participants were instructed to select those which were true. If the participant selected any options which were incorrect, their survey ended, they did not go on to the following sections, and their data was discarded.

If they selected the correct options, participants moved on to the takeaway section of the survey. In this section, participants were shown a chart and told that they would be asked to report takeaways on the following page. They were also informed that they would not be able to go back and view the chart after moving on. They selected a button to indicate they were ready to move forward. 

On the page following the image of the chart, participants were asked to provide their takeaways. They were required to provide at least one but had the option to provide up to three takeaways.  They were also asked to rate ``Which aspect of the information did you rely on when listing your takeaways?'' on a scale from 1 (``entirely text'') to 5 (``entirely visual''). 

The fourth section of this survey was a subjective measure of graphical literacy \cite{garcia2016measuring}. Garcia-Retamero et al. tested this subjective assessment of graphical literacy and compared its performance to an objective measure \cite{galesic2011graph}, finding similar and substantial predictive power. The measure also assesses preference, which is a key and central question of this paper. The measure has subsequently been used in other studies as a measure of graphical literacy \cite{millet2021end, yang2021explaining}.

Within this measure, participants were asked to report how good they considered themselves to be at working with different chart types, such as line, bar, and pie charts. They were also asked questions regarding their preference for textual or visual information, such as ``When reading books or newspapers, how helpful do you find charts that are part of a story?'' and ``To what extent do you believe in the saying `a picture is worth one thousand words'?''. We added an additional question to this section, ``To what extent do you believe in the saying, ‘Reading expands the mind'?'' to add a text-focused counterpart to the ``thousand words'' image-focused question.

Finally, participants completed the demographic section. Here, they reported their age range (e.g., 18-24), their current education level (e.g., ``Less than high school''), and their experience with charts and reading. This experience was captured through both the frequency (e.g., ``Every day'') as well as the context in which they engaged with the material (e.g., Government reports). For reading, we inquired about the frequency of both short-form reading (e.g., text messages, tweets) as well as long-form (e.g., books, magazines). 

\section{Results}
\label{section:results}

In this section, we provide results for research questions RQ1 -- RQ3.  Each research question is assessed in terms of a set of hypotheses as enumerated in Section \ref{section:study}.
Analysis plans were preregistered on OSF.\footnote{\url{https://osf.io/vz976/registrations}} 

\pheading{Visual vs. Textual Preferences.}
Hypotheses below compare participants with an overall preference for visual information (the \textsc{visual} group) and those with an overall preference for textual information (the \textsc{textual} group). To calculate a preference score for each participant, we summed the Likert ratings for questions included in the measure of graphical literacy \cite{garcia2016measuring} as well as questions regarding the frequency of reading text and viewing charts included in the demographics section.

Participants with scores in the top 25\% of the set were placed in the \textsc{visual} group, and those with scores in the bottom 25\% were placed in the \textsc{textual} group. Because we were primarily interested in participants with a distinct preference for visual or textual information, we did not place the middle 50\% of the set into either group. These participants were not examined for the hypotheses that compared preferences.

\subsection{RQ1: Preferences for Amount of Text}
RQ1 asks: what are readers' preferences when viewing charts with different amounts of text?

\pheading{Summary of Findings:} Overall, we found a strong preference for the charts with a greater number of annotations. For both sets ranked, the highest performing stimulus was the chart with the most annotations. 
They 
rated the no-text variant or title-only variant consistently lowest.
Further discussion on preferences for other variants can be found in Section \ref{section:discussion}.

\begin{figure*}[t]
 \centering
 \includegraphics[width=0.9\linewidth]{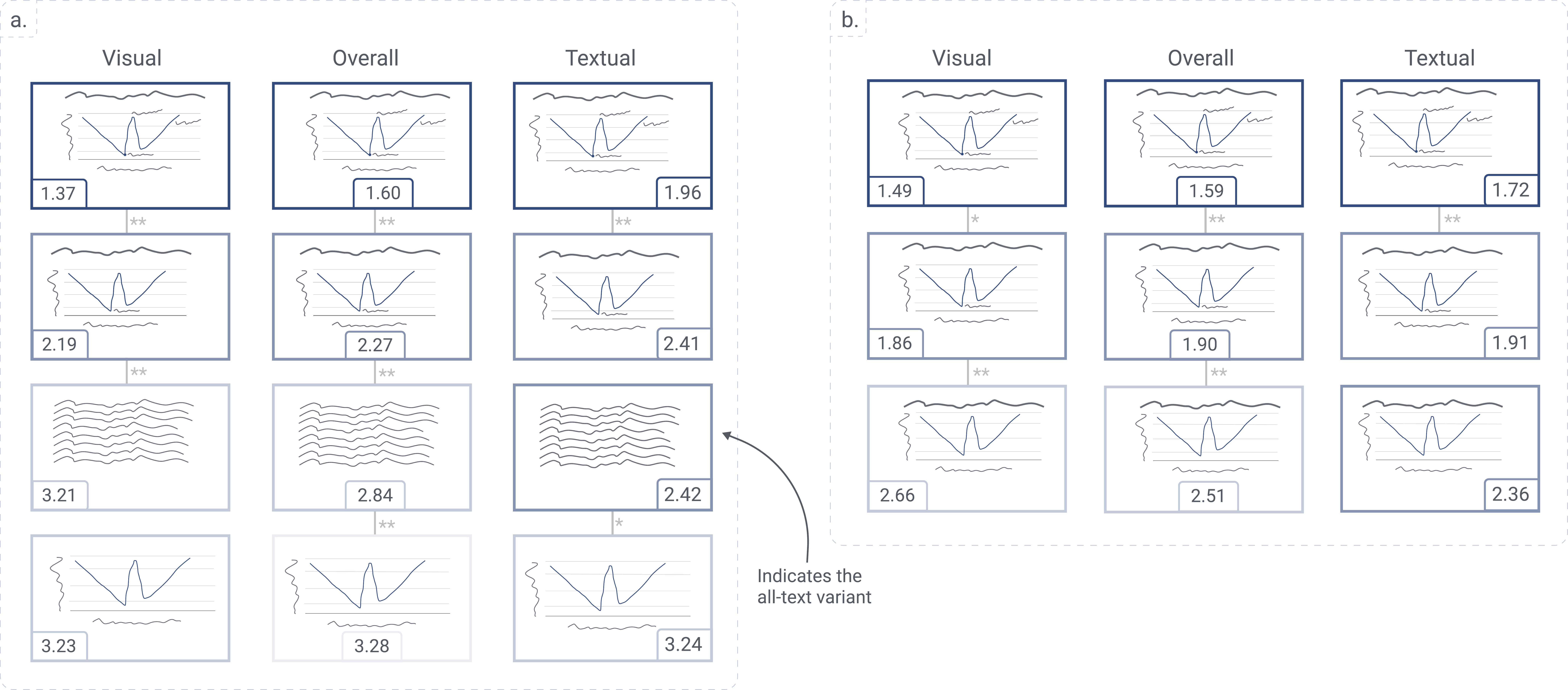}
 \caption{Ranking task results, shown for all participants (Overall) as well as the \textsc{visual} and \textsc{textual} groups.
 Pairwise significance can be seen in Tables \ref{tab:rank_setA} and \ref{tab:rank_setB}. Lines within columns indicate statistically significant differences between average rankings (* = 0.05, ** = 0.01). 
 \newline (a) Ranking set A (`extremes'). Possible values range from 1 (ranked first) to 4 (ranked last).
 (b) Ranking set B (`fine-grained'). Possible values range from 1 (ranked first) to 3 (ranked last).
 }
 \label{fig:ranking_results}
\end{figure*}

\pheading{Details of Analysis:} To examine this question more closely, we completed Friedman's tests for the rank-order data. We performed post-hoc Nemenyi's tests to examine pairwise comparisons \cite{pereira2015overview}, shown in Tables \ref{tab:rank_setA} and \ref{tab:rank_setB}. Overall relative stimuli rankings can be seen in Figure \ref{fig:ranking_results}.

In both ranking tasks, the charts with the most annotations were ranked highest overall. In Ranking Set A, the annotation+ variant was highest ranked across \textsc{visual} and \textsc{textual} participant groups. For the \textsc{textual} group, title-annotation1 did not differ from the textual paragraph. The sparse charts with minimal annotations ranked lowest across conditions -- participants least preferred the no-text variant from Set A and the title-only variant from Set B.  

Hypotheses 1a and 1b refer to the rankings of the `extremes', shown in Figure \ref{fig:teaser}, while 1c and 1d refer to the `fine-grained' rankings, shown in Figure \ref{fig:ranking_sets}. 

\subsubsection{H1a, b: Preferences Compared to All-Text}

\textbf{H1a:} \textit{Readers with an overall preference for visual information
will prefer charts with annotations compared to the all-text variant.} 

This hypothesis was \textbf{supported} ($\chi^2 = 151.15$, df = 3, $p < 0.01$). For the \textsc{visual} group, annotation+ variant received the highest average rating (1.37), while the all-text and no-text variants received the lowest rating (3.21, 3.23).

\smallskip
\noindent 
\textbf{H1b:} \textit{Readers with an overall preference for textual information will prefer the all-text variant compared to charts with little or no text.}

This hypothesis was \textbf{partially supported}. Overall, the average rankings for the stimuli differed ($\chi^2 = 50.962$, df = 3, $p < 0.01$). For the \textsc{textual} group, the all-text variant (2.42) did not significantly differ from the title-annotation1 variant (2.41). Therefore, there was no significant difference between all-text and a chart with little text. However, the no-text chart was rated lowest (3.24). So, we find partial support for this hypothesis in comparison to the no-text.

Although most participants ranked the annotation+ chart in Ranking Set A highest, data also show that a substantial minority\,---\,42, or 14\% of participants\,---\,ranked the all-text option as their first choice for Set A.  Of these, 25 are identified by our analysis as \textsc{textual}, 5 as \textsc{visual} preferred.  
One author examined the comments participants made when discussing the all-text variant.
Seven of these 42 participants indicated that they had a general preference for text over charts. Other reasons included the text being perceived as clearer, more understandable, and/or providing a better explanation (13), providing a better story or more interesting context (4), or more precise, concrete, or thorough (9). A few (15) participants ranked the no-text chart highest. The most common reasons given were simple / easy to understand (9), clean/lack of clutter (5), and allowing the viewer to draw their own conclusions (4).  

\subsubsection{H1c, d: Preferences For Quantity of Annotations}

\smallskip
 \noindent 
\textbf{H1c}: \textit{Readers with an overall preference for visual information will prefer charts with fewer annotations.}

This hypothesis was \textbf{not supported}, though the average rankings of stimuli did differ ($\chi^2 = 75.25$, df = 2, $p<0.01$). The \textsc{visual} group most preferred charts with the greater number of annotations, rather than those with fewer.

\smallskip
 \noindent 
\textbf{H1d}: \textit{Readers with an overall preference for textual information will prefer charts with a greater number of annotations.}

This hypothesis was \textbf{supported}, as results were similar to the \textsc{visual} group ($\chi^2 = 21.24$, df = 2, $p<0.01$). The \textsc{textual} group also most preferred charts with the greater number of annotations, rather than those with fewer.

\begin{table}[ht]
\def\arraystretch{1.5}
   \centering
     \caption{Detailed data about takeaways and viewed annotations, by semantic level. The rows, from top to bottom, show the total number of takeaways produced in each level, the number of annotations viewed for each level, the number of takeaways produced in each level after viewing an annotation in the given level, and the number of takeaways which matched the annotations at each level.}
     \begin{tabular}{ | m{0.3\linewidth} | m{0.11\linewidth} | m{0.11\linewidth} | m{0.11\linewidth} | m{0.11\linewidth} | } 
      \cline{2-5}
      \multicolumn{1}{c|}{}  & L1 & L2 & L3 & L4\\ 
      \hline
      Total takeaways & 48 & 132 & 446 & 110 \\
      \hline
      Annotations viewed & 284 & 304 & 316 & 287 \\
      \hline
      Takeaways after \newline viewing annotations & 23 (8\%) & 73 (24\%) & 316 (65\%) & 100 (35\%) \\
      \hline
      Matched takeaways & 17 (29\%) & 60 (32\%) & 80 (22\%) & 88 (81\%) \\
      \hline
  \end{tabular}

    \label{tab:takeaway_stats}
\end{table}

\begin{table*}[t]
\def\arraystretch{1.5}
    \centering
    \caption{Pairwise comparisons between preference groups for the 'extremes' ranking set (A). Annotation abbreviated to `annot'.}
    \begin{tabular}{ | m{0.12\linewidth} | m{0.12\linewidth} | m{0.12\linewidth} | m{0.12\linewidth} | m{0.12\linewidth} | m{0.12\linewidth} | m{0.12\linewidth} | m{0.12\linewidth} |} 
      \cline{2-7}
      \multicolumn{1}{c|}{}
        & No-text, \newline Title-annot1 & No-text, \newline  Annot+ & No-text, \newline All-text & Title-annot1, \newline   Annot+ & Title-annot1, \newline All-text &  Annot+, All-text \\ 
      \hline
      Visual & \cellcolor{blue!15}p $<$ 0.01 & \cellcolor{blue!15}p $<$ 0.01 & p = 1 & \cellcolor{blue!15}p $<$ 0.01 & \cellcolor{blue!15}p $<$ 0.01 & \cellcolor{blue!15}p $<$ 0.01 \\ 
      \hline
      Textual & \cellcolor{blue!15}p $<$ 0.01 & \cellcolor{blue!15}p $<$ 0.01 & \cellcolor{blue!15}p $<$ 0.01 & \cellcolor{blue!5}p $=$ 0.48 & p = 1 & \cellcolor{blue!5}p $=$ 0.041 \\ 
      \hline
     Overall & \cellcolor{blue!15}p $<$ 0.01 & \cellcolor{blue!15}p $<$ 0.01 & \cellcolor{blue!15}p $<$ 0.01 & \cellcolor{blue!15}p $<$ 0.01 &\cellcolor{blue!15}p $<$ 0.01 & \cellcolor{blue!15}p $<$ 0.01 \\ 
      \hline
    \end{tabular}
    \label{tab:rank_setA}
\end{table*}

\begin{table}[ht]
\def\arraystretch{1.5}
   \centering
   \caption{Pairwise comparisons between preference groups for the 'fine-grained' ranking set (B). Annotation abbreviated to `annot'.}
    \begin{tabular}{ | m{0.11\linewidth} | m{0.23\linewidth} | m{0.23\linewidth} | m{0.23\linewidth} |} 
      \cline{2-4}
      \multicolumn{1}{c|}{}  & Title-only, \newline Title-annot1 & Title-only, \newline Title-annot2 & Title-annot1, \newline Title+two \\ 
      \hline
      Visual & \cellcolor{blue!15}p $<$ 0.01 & \cellcolor{blue!15}p $<$ 0.01 & \cellcolor{blue!5}p $=$ 0.028 \\ 
      \hline
      Textual & \cellcolor{blue!15}p $<$ 0.01 & \cellcolor{blue!15}p $<$ 0.01 & p = 0.352 \\ 
      \hline
      Overall & \cellcolor{blue!15}p $<$ 0.01 & \cellcolor{blue!15}p $<$ 0.01 & \cellcolor{blue!15}p $<$ 0.01 \\ 
      \hline
    \end{tabular}
    \label{tab:rank_setB}
\end{table}

\subsection{RQ2: Semantic Levels and Takeaways}

RQ2 asks: How do the different semantic levels of text affect the type of information included in readers' takeaways? Further, when do readers include information in their takeaways that are not found either in the chart or the text? Which semantic levels of text do readers rely on when they form their takeaways? 

\pheading{Summary of Findings:} Through the following analyses, we found that participants' takeaways were overall influenced by the annotations present and influenced by \textit{some} semantic levels more than others. Specifically, annotations belonging to L2 and L4 had the greatest influence on the participants' takeaways, followed by L3. L1 annotations had almost no effect, in part because participants made very few L1 takeaways.

Overall, participants were most likely to provide L3 takeaways. When provided with an L4 annotation in the chart, participants were more likely to include their own knowledge or experience, though they rarely did this. The semantic level of the annotation text did not affect the degree to which participants relied upon the text.

\pheading{Details of Analysis:} Prior to investigating the hypotheses, two authors independently coded participant takeaways for the semantic level of the takeaway, whether the takeaway was a match to any of the provided annotations in the chart viewed, and whether the participant incorporated any information not contained in the chart and the text provided (e.g., a real-world event). In order to be considered a ``match'', the takeaway should contain exact or similar language as used in the annotation provided. If a takeaway contained language belonging to more than one semantic level, the higher semantic level ``trumped'' the lower level. The takeaway was then assigned the higher level (e.g., converting L2 to L3). This method follows the process used by Lundgard \& Satyanarayan \cite{lundgard2021accessible}.

Overall, inter-rater reliability was 0.818 (Kappa = 0.815 for takeaway level, Maxwell's RE = 0.727 for match, and Maxwell's RE = 0.912 for external information) \cite{gamer2012package}. A third author coded the conflicts and resolved all but four, which were then discussed amongst the coders.

Overall, participants provided 736 takeaways. On average, each participant wrote 1.83 takeaways. We did not instruct participants to provide takeaways in any particular order and did not treat takeaways differently depending on how many the participant provided.

Participants provided 48 takeaways classified as L1, 132 classified as L2, 446 classified as L3, and 110 classified as L4. 245 of the 736 total takeaways (33.3\%) matched the annotations provided in the chart. Further information on the takeaway levels can be found in Table \ref{tab:takeaway_stats}.

\subsubsection{H2a: Semantic Alignment between Text and Takeaways}

\textbf{H2a:} \textit{Reader takeaways will be more likely to contain a given semantic level if the text accompanying the chart also contains that semantic level. This will be true for all semantic levels.}

\smallskip
To examine 2a, we used logistic regression to analyze the relationship between the presence of different semantic levels of text on the chart and the semantic level of the reader's takeaway. We found \textbf{partial support} for this hypothesis.

An example of an L1 takeaway is, ``\textit{The date range was 1960 to 2020}," (P68). As mentioned previously, L1 takeaways were relatively uncommon. There were no significant influences on what semantic level of annotation may predict an L1 takeaway. So, H2a was not supported for L1.

An example of an L2 takeaway is, ``\textit{Sports game attendance was lower than 5000 around 2009}," (P75). Participants were more likely to make takeaways containing L2 when they had viewed a chart that provided an L2 annotation, compared to all other semantic levels. They were \underline{\smash{1.6 times as (hereafter 1.6x) likely}} to provide an L2 takeaway when they were provided an L2 annotation than when they were shown an L1 annotation (95\% CI: [1.06, 2.40], $p < 0.05$),  \underline{\smash{2.0x likely}} compared to L3 ([1.29, 2.95], $p < 0.01$), and \underline{\smash{1.8x likely}} compared to L4 ([1.44, 2.22], $p < 0.01$). In summary, L2 takeaways were most likely to occur when the chart itself included L2 annotations. Therefore, H2a is supported for L2.

An example of an L3 takeaway is, `\textit{The national debt fell steeply in the early 90s before quickly rising to an all-time high}," (P4). L3 takeaways were the most frequently provided by participants overall. Participants were \underline{\smash{2.2x as likely}} to provide an L3 takeaway when they were provided L3, in comparison to when they were provided L4 ([1.59, 3.05], p $<$ 0.01). Comparisons to other annotation levels did not affect the likelihood of making an L3 takeaway. Specifically, after seeing an L3 annotation, participants were not more likely to make an L3 takeaway compared to when they saw an L1 or L2 annotation.

Overall, participants provided mostly L3 takeaways. H2a is partially supported for L3, as the presence of an L3 annotation made it more likely that the participant would make an L3 takeaway, but not in comparison to all semantic levels.

An example of an L4 takeaway is, ``\textit{The tech boom in the early 1980s lead to a huge increase in median household incomes}," (P215). Participants were significantly more likely to make L4 conclusions when they were provided with an L4 annotation: \underline{\smash{6.1x as likely}} than when provided with L1 ([3.82, 10.19], $p < 0.01$); \underline{\smash{6.2x as likely}} than when provided with L2 ([3.85, 10.11], $p < 0.01$); \underline{\smash{3.2x as likely}} than when provided with L3 ([2.16, 4.79], $p < 0.01$).  L4 takeaways were most likely to occur when the chart itself included L4 annotations. Therefore, H2a is supported for L4. 

\subsubsection{H2b: L1  Content and External Takeaways}

\textbf{H2b:} \textit{Readers will be more likely to include information in their takeaways that are not found in the visual information or text when L1 is present in the chart.}

\smallskip
To examine 2b, we used logistic regression to analyze the relationship between the presence of different semantic levels of text on the chart and the existence of external information in the participant's takeaway. This hypothesis was \textbf{not supported}. Readers were more likely to include external information when L4 was present in the chart, compared to L1 ([1.15, 72.73], p $<$ 0.05). However, there were very few takeaways that incorporated external information, as indicated by the extremely wide confidence interval. It remains unclear what kinds of annotations may encourage readers to include external information in their takeaways, but it was done infrequently in this case.

\subsubsection{H2c, d, e: Preferences and Reliance on Visual vs. Text}

\noindent \textbf{H2c}: \textit{Readers will self-report as relying less on text when the text contains L1.}

\smallskip
\noindent \textbf{H2d}: \textit{Readers' will self-report as relying more on text when the text contains L4.}

\smallskip
\noindent \textbf{H2e}: \textit{Readers with an overall preference for textual information will rely more on text within a chart.}

\smallskip
To examine hypotheses 2c-e, we performed a one-way ANOVA with post-hoc testing with Bonferroni correction. Reliance on textual and visual components of the chart was reported by participants on a 5-point Likert scale, where 1 indicated complete reliance on text and 5 indicated complete reliance on visual components. 

Hypotheses 2c-d were \textbf{not supported}. Although there were differences between self-reported reliance on text between semantic levels ($F = 5.81$, $p = 0.016$, df = 1), follow-up pairwise testings with Bonferroni correction were not significant ($p > 0.05$). Average reliance ratings can be seen in Table \ref{tab:reliance_levels}. 

\begin{table}[ht]
\def\arraystretch{1.5}
    \centering
     \caption{Average reliance scores (1 = `entirely text', 5 = `entirely visual') between semantic levels.}
     \begin{tabular}{ | m{0.24\linewidth} | m{0.13\linewidth} | m{0.13\linewidth} | m{0.13\linewidth} | m{0.13\linewidth} |} 
      \hline
        \textbf{Semantic Level} & L1 & L2 & L3 & L4 \\ 
      \hline
       \textbf{Avg. Reliance} & 3.84 & 3.76  & 3.65 & 3.60 \\ 
      \hline
    \end{tabular}
    \label{tab:reliance_levels}
\end{table}

There was a difference between the \textsc{textual} and \textsc{visual} groups; H2e was \textbf{supported}. The \textsc{textual} group reported they relied more on the text in the chart (3.53) than the \textsc{visual} group (3.87, Bonferroni-adjusted p $=$ 0.012).

\subsection{RQ3: Effect of Text Placement on Takeaways}

RQ3 asks: How does the placement of text affect how the reader integrates the text with the chart? 

\smallskip
\pheading{Summary of Findings:} Overall, the role of the placement of text varied between semantic levels. Participants were most likely to match the L3 annotation provided when it was in the `title' position, but this effect was not true for any other level. For the other three levels, participants were more likely to match the annotation when the text was closer in proximity to the chart. For L1, this meant being anchored to one of the chart axes. For both L2 and L4, this meant being anchored to a specific point on the chart, accompanied by a visual annotation of the point. 

\pheading{Details of Analysis:} The method for qualitatively coding the `position' of an annotation varied from that which was preregistered. In the preregistration, the chart was divided into quadrants (e.g., top left), and the annotations were assigned a position based on their physical position on the chart. However, the precise physical position is not as relevant to this paper as the semantic position (e.g., as a title). Therefore, we analyze the positions coded as follows.

Each annotation was assigned a `position' during the stimuli creation stage to ensure counterbalancing of the title position. For L1, other possible positions included `axis', with the annotation in a gray box next to the x- or y-axis, as well as `trend', for which the annotation was positioned in an unanchored callout box. For L2--L4, positions also included `trend,' in which the text was positioned next to an arrow or a gray band highlighting the relevant years. These levels were also positioned as a `point', which marked the point of interest with a dot matching the line in color. 
\subsubsection{H3: Position and Match of Takeaway}
\noindent \textbf{H3}: \textit{Readers' takeaways will be most likely to match a given semantic level if the text containing that semantic level is positioned as a title.}

\smallskip
To examine H3, we used logistic regression to analyze the relationship between the position of different semantic levels of text on the chart and whether the reader's takeaway was a match to the annotation provided in the chart. We found \textbf{partial support} for this hypothesis.

For takeaways that matched the L1 annotations, participants were most likely to match their takeaways to the annotation when it was positioned by an axis in the chart rather than as a title. In fact, they were \underline{\smash{11.3x as likely}} to match the annotation when it was positioned as an axis than when it was a title ([3.18, 40.01], $p < 0.01$). So, this hypothesis was not supported for L1 takeaways. Titles were matched less often, while axis comments were matched most often.

Participants were most likely to match the L2 annotations when they were positioned as points rather than titles. They were \underline{\smash{2.0x as likely}} to match the annotation as a point than as a title ([1.00, 3.98], $p = 0.485$). Thus, H3 was also not supported for L2 takeaways.

L3 takeaways, on the other hand, were most likely to be matched by participant takeaways when they were positioned as the title. They were \underline{\smash{0.49x as likely}} to be matched when positioned as a point compared to positioned as a title ([0.380, 0.620], $p < 0.01$) and \underline{\smash{0.38x as likely}} when comparing annotations positioned as a trend to those as a title ([0.293, 0.506], $p < 0.01$). H3 was supported for matched L3 annotations. 

In continuation of the overall trend, L4 takeaways were most likely to match when they were positioned as points, rather than as titles. Participants were \underline{\smash{1.9x as likely}} to match the L4 annotation provided by the chart when it was positioned in the role of a point, compared to that of a title ([1.11, 3.22], $p = 0.018$). 
\section{Discussion}
\label{section:discussion}

This paper explores the effects of text placement and the various semantic levels on readers' preferences and takeaways. We discuss findings from our study and identify themes that support a better understanding of chart and text integration, along with practical guidelines for visualization tools and authoring practices. We use the notation $P\#$ when referring to participants in the discussion.

\subsection{Preferences for annotation}

General guidelines for information sharing and presentation have shied away from adding too much text to imagery at the risk of overwhelming the viewer or not conforming to good design practices \cite{tufte1985visual}. However, our study shows that more annotation was not penalized and was, in fact, often preferred. Readers liked the blend of text with charts, ranking charts annotated with more text higher than their sparser counterparts. P10 indicated that, although they found the paragraph of text helpful, \textit{``the visuals of the graph + text allow me to more quickly assess the information.''} P31 called the combination \textit{``the best of both worlds.''}

Charts with the most annotations were especially preferred by those with preferences for visual information. P158 commented, ``\textit{I think all the extra facts and text on the chart are excellent ways to help people really know what's going on in the chart.}'' P45 expressed that, ``\textit{the chart that combines both visual and written information... helps the widest variety of learning styles to understand the chart.}'' 

This result runs counter to concerns about the potential for such annotations to make the visual components of the chart appear cluttered. Many participants noted that they preferred the charts with more annotations despite the fact that they might appear more cluttered.

\subsection{Preferences for text}

Additionally, a non-trivial minority (14\%) of participants preferred a textual paragraph over a visual chart. Readers found the text ``\textit{Easier to understand... when written out rather than the busy graph}'' [P15]. Text was also a conducive medium for adding external information to support the data narrative - ``\textit{Like seeing the different points explained thoroughly and written in a story-like format}'' [P41]. 

Although charts with annotations were generally preferred, the \textsc{textual} group ranked the text-only variant higher than the chart-only variant, which we did not observe in the \textsc{visual} group. P178, who preferred the text-only variant, reported that they ``\textit{like to read the words and use my own imagination for the imagery.}'' P6 called the annotation+ chart, ``\textit{ugly to look at,}'' a sentiment which was echoed by many who ranked that variant low. 

Others highlighted the distinction in preferences between \textsc{textual} and \textsc{visual} groups - ``\textit{I like this because it's very clear and gives me the information I need to know. I'm content with this. A more visual learner, however, would probably prefer the graph,}'' [P139]. This feedback adds support to findings from prior work and underscores the importance for visualization research to consider the text-only case when comparing visualization options \cite{hearst2019would, stokesgive}.

\subsection{Semantics and positions}

We found that the semantic levels of text included with the chart can influence the kind of information that readers include in their takeaways. We observed that readers
were more influenced by annotations belonging to L2 or L4.

Including external context to add to the narrative of the chart inspires readers to include that information in their own takeaways, but some participants would prefer to draw their own conclusions about the data, similar to Lundgard \& Satyanarayan \cite{lundgard2021accessible}. For instance, P258 said ``\textit{I like the raw data and being able to draw my own conclusions from the data. Data doesn't lie.}'' about their preference for the no-text variant.

We also found that the proximity of text to what it is describing in the chart influences reader takeaways. For instance, readers included L1-text in their takeaways when the L1-text was describing one of the axes as an on-chart annotation. L2- and L4-text placed near a data point on the chart that they described, were more likely to be included in the takeaway as opposed to being positioned as the title of the chart. On the other hand, readers described overall trends from L3-text when it was placed as a title in the chart. These observations highlight an interesting interplay between spatial and text semantics in charts.

\subsection{Visual language and aesthetics}

In addition to adding text annotations, we employed a visual language consisting of shaded boxes, dots, and arrows. Some readers responded positively to this visual language, ``\textit{I like the grey arrows shown next to text explaining why they are pointing in the direction they are,}'' [P19] or even requesting that an arrow be added to a chart that lacked it [P136]. However, another participant said these features ``\textit{provide unnecessary visual clutter,}'' [P223]. Many readers also requested the use of color, improving the chart ``\textit{by livening it up a bit,}'' (P31) or by making it ``\textit{more eye-catching,}'' [P112]. 

Another area of conflict between reader opinions was around the simplicity of the chart. Some enjoyed the simple nature of the visualization, while others told us to make it ``\textit{more interesting (think Edward Tufte),}'' [P28]. Readers expressed opinions about the positioning or length of text when the chart lacked a title or when the title was too long: ``\textit{used to pretty short and concise titles,}'' [P135]. 

\subsection{Design Guidelines and Implications}

Designers often add text such as captions along with charts to provide additional context, and previous work indicates the benefits of having both text and charts emphasize the same information \cite{kim2021towards}. However, the amount, content, and positioning of these textual components can vary widely. From the findings described in Section \ref{section:results}, we provide a set of guidelines for univariate line charts. These guidelines offer insight into what type of content to express and where to place text on charts to elicit desired takeaways. Future work in more complex or other chart types can serve to expand these guidelines.

\smallskip
\noindent
\textbf{Guideline 1: Rather than aiming for maximally minimalist design, annotate charts with relevant text.} \textit{(H1a-d)}

Appropriately designed univariate charts that contain multiple textual annotations are preferred by a majority of visually-oriented and textually-oriented readers. Higher quantities of annotations are not penalized for the `clutter' they may cause. Rather, the provided information is seen to enhance the view of the data.

\smallskip
\noindent
\textbf{Guideline 2: The best semantic content depends on the intended message.} \textit{(H2a)}

The phrasing of the text can nudge readers towards certain takeaways. Depicting statistical information as textual annotations, such as identifying a maximum value or comparing two points, makes readers more likely to take away this type of information. By contrast, describing the chart’s axes and encodings does not elicit takeaways of this type of information. Including external information, such as current events or sociopolitical context,  makes the reader more likely to take away this type of information. Readers are likely to make takeaways concerning perceptual components of the chart; annotations with perceptual language, such as trends and peaks, may not have an effect. 

\smallskip
\noindent
\textbf{Guideline 3: To convey a message, the best placement of the text depends on the semantic content of the text.} \textit{(H3)}

Content describing chart axes and encodings should be positioned by a relevant axis. Statistical or numerical information should be positioned by a relevant point. Perceptual content relating to patterns or trends should be positioned in the title. Any external information or context should be positioned by a relevant point.

\smallskip
\noindent
\textbf{Guideline 4:  Consider a text-only variant that can stand alone.} \textit{(H1a-b)}

A subset of the population prefers text-only over text alongside a visualization; when designing a presentation or performing studies of different visualizations, consider a version of the text that can stand on its own. This text can also be a useful starting point for story-centered textual annotation and provide accessibility benefits \cite{jung2021communicating}.

\section{Limitations and Future Work}
\label{section:lfw}

The study results highlight the importance of text with charts and how they together influence how readers glean insights from the data. The results also point out that reader preferences are critical in determining the full experience. These observations provide theoretical and practical implications for a better understanding of text in the larger context of visual analysis.

In this work, we explore the interplay between text and univariate line charts. Limiting some of the design options was necessary since  the combinatorial design space that accounts for the various semantic levels, amount, and placement of text, is  large and complex to evaluate. We controlled for the amount of text in the charts by varying the presence or absence of one or more semantic levels of text. It would be useful to create  a more systematic way of determining text progression, with a more consistent computation of the amount of text and visual information in charts. Future work should explore how readers integrate text with other chart types, including dashboards and other contexts such as mobile devices, print medium, and chatbot interfaces. 

We employed synthetically generated line charts to control for the experimental factors but manually crafted the placement of annotations and embellishments such as arrows with the guidance of a visualization expert. The specifics of visual annotations are also a wide and varied design space that future work can explore. The line charts created clearly showed abrupt peaks and trends, which is not often the case in practice or in real examples. We also used fictional examples to decrease the influence of prior knowledge on a topic.

While the annotation process provided some ecological validity, we would need to explore other design parameters (e.g., color) and include real-world examples to better understand other influences on readers' preferences and takeaways. The absences of a text document embedding  may have introduced a potential bias towards textual representations.

Additionally, the all-text variant naturally included more L4 content than some of the charts. This content was most preferred by sighted readers and may have provided an advantage to the all-text variant \cite{lundgard2021accessible}. The annotation+ chart variant also included L4 information, but the annotations were briefer than the full sentences in the paragraph. The text in the stimuli was worded in a neutral manner; results may differ if the text were worded in a manner perceived as biased.

For our analysis, we used one part of a graphical literacy measure \cite{garcia2016measuring} that was correlated with preference, combining these responses with participants' self-reported behaviors around viewing charts and reading in long- and short-form. However, this analysis may not provide clear or rigorous insight into the overall preference a reader has toward charts or text. Future work should develop and employ a more nuanced measure to capture and compare groups with different preferences.

\section{Conclusion}
\label{section:conclusions}

In this paper, we conducted a set of studies on nine examples of univariate line charts with varied amounts and types of annotations to better understand reader preferences and takeaways from those charts. Findings show that readers prefer a combination of textual and visual communication over a single medium. They also provide further evidence of preference differences and indicate that the choice of semantic content and placement of the text in a chart can influence the readers’ takeaways. We propose guidelines to better design charts with text; by using placement and appropriate semantic content, visual features in the chart can be better supported by their descriptions in the text. 

\acknowledgments{
This research was supported in part by a gift from the Allen Institute for AI and NSF award \#1900991.
}

\bibliographystyle{abbrv-doi}

\newpage

\bibliography{references}
\end{document}